\documentclass[aps,pre,floatfix,reprint,longbibliography]{revtex4-2}
\usepackage[top=2cm,bottom=2cm,left=1.5cm,right=1.5cm,columnsep=0.7cm]{geometry}
\usepackage{xparse,xcoffins}
\usepackage{float}
\mathchardef\mhyphen="2D

\ExplSyntaxOn
\NewCoffin\imagecoffin
\NewCoffin\labelcoffin

\keys_define:nn { alex/label }
 {
  label   .tl_set:N = \l_alex_label_tl,
  labelbox .bool_set:N = \l_alex_label_box_bool,
  labelbox .default:n = true,
  fontsize .tl_set:N = \l_alex_label_size_tl,
  fontsize .initial:n = \footnotesize,
  pos .choice:,
  pos/nw .code:n = \tl_set:Nn \l_alex_label_pos_tl { left,up },
  pos/ne .code:n = \tl_set:Nn \l_alex_label_pos_tl { right,up },
  pos/sw .code:n = \tl_set:Nn \l_alex_label_pos_tl { left,down },
  pos/se .code:n = \tl_set:Nn \l_alex_label_pos_tl { right,down },
  pos/n .code:n = \tl_set:Nn \l_alex_label_pos_tl { hc,up },
  pos/w .code:n = \tl_set:Nn \l_alex_label_pos_tl { left,vc },
  pos/s .code:n = \tl_set:Nn \l_alex_label_pos_tl { hc,down },
  pos/e .code:n = \tl_set:Nn \l_alex_label_pos_tl { right,vc },
  pos .initial:n = nw,
  unknown .code:n   = \clist_put_right:Nx \l_alex_label_clist
                       { \l_keys_key_tl = \exp_not:n { #1 } }
 }
\clist_new:N \l_alex_label_clist
\box_new:N \l_alex_label_box
\box_new:N \l_alex_label_image_box

\NewDocumentCommand{\xincludegraphics}{O{}m}
 {
  \group_begin:
  \tl_clear:N \l_alex_label_tl
  \clist_clear:N \l_alex_label_clist
  \keys_set:nn { alex/label } { #1 }
  \tl_if_empty:NTF \l_alex_label_tl
   {
    \alex_includegraphics:Vn \l_alex_label_clist { #2 }
   }
   {
    \SetHorizontalCoffin\imagecoffin
     {
      \alex_includegraphics:Vn \l_alex_label_clist { #2 }
     }
    \SetHorizontalCoffin\labelcoffin
     {
      \raisebox{\depth}
       {
        \bool_if:NTF \l_alex_label_box_bool
         { \fcolorbox{white}{white}{\l_alex_label_size_tl\l_alex_label_tl} }
         { \l_alex_label_size_tl\l_alex_label_tl }
       }
     }
    \SetVerticalPole\imagecoffin{left}{35pt+\CoffinWidth\labelcoffin/2}
    \SetVerticalPole\imagecoffin{right}{\Width-3pt-\CoffinWidth\labelcoffin/2}
    \SetHorizontalPole\imagecoffin{up}{\Height--11.5pt-\CoffinHeight\labelcoffin/2}
    \SetHorizontalPole\imagecoffin{down}{3pt+\CoffinHeight\labelcoffin/2}
    \use:x{\JoinCoffins\imagecoffin[\l_alex_label_pos_tl]\labelcoffin[vc,hc]} 
    \TypesetCoffin\imagecoffin
   }
   \group_end:
 }
\NewDocumentCommand{\setlabel}{m}
 {
  \keys_set:nn { alex/label } { #1 }
 }

\cs_new_protected:Nn \alex_includegraphics:nn
 {
  \includegraphics[#1]{#2}
 }
\cs_generate_variant:Nn \alex_includegraphics:nn { V }

\ExplSyntaxOff

\usepackage{bbm} 

\usepackage[T1]{fontenc}
\usepackage[utf8]{inputenc}

\usepackage{amsfonts,amssymb,amsmath,array}
\usepackage{bm}
\usepackage{dcolumn}
\usepackage[output-decimal-marker={.}]{siunitx} 

\usepackage[caption=false]{subfig}
\usepackage{comment}

\usepackage{graphicx} 
\usepackage{listings} 
\usepackage{microtype}
\usepackage{color}
\usepackage{float}

\usepackage{hyperref}
\hypersetup{
    colorlinks=true,
    linkcolor={blue}, 
    filecolor=magenta,   
    citecolor={blue}, 
    urlcolor=blue, 
}

\newcommand{\br}{{\bf r}}

\newcommand{\bF}{{\bf F}}
\newcommand{\bn}{{\bf n}}

\usepackage{enumitem}



\begin{document}

\preprint{APS/123-QED}

\title{Proliferating active disks with game dynamical interaction
}

\author{Alejandro Almod\'ovar}
\email{almodovar@ifisc.uib-csic.es}
\author{Tobias Galla}
\email{tobias.galla@ifisc.uib-csic.es}
\author{Crist\'obal L\'opez}
\email{clopez@ifisc.uib-csic.es}
\affiliation{IFISC, Instituto de F\'isica Interdisciplinar
y Sistemas Complejos (CSIC-UIB), Campus Universitat de les
Illes Balears, E-07122 Palma de Mallorca, Spain}

\begin{abstract}

We study a system of self-propelled, proliferating finite-size disks with game-theoretical interactions, where growth rates depend on local population composition. We analyze how these interactions influence spatial distribution, coexistence, and extinction. Three scenarios emerge: (i) stable coexistence with well-mixed distributions, (ii) bistability, where the outcome depends on initial conditions, and (iii) dominance, where one species always outcompetes the other. We further show that activity enhances spatial mixing in coexistence and reduces extinction times in both dominance and bistability. Additionally, stronger interactions promote coexistence while accelerating extinction in the latter two regimes. By varying diffusivity, activity, and interaction strength, we show how movement and spatial constraints affect population dynamics.

\end{abstract}

\maketitle

\section{Introduction}
\label{Sec:intro}

Various biological processes,
such as tissue formation \cite{Angelini2011},
wound healing \cite{XTrepat2020, XTrepat2021, Sepulveda}, microbial biofilm growth \cite{Szabo2006}, and
tumor development \cite{Zapperi}, involve the interaction 
of motile individuals that proliferate and/or die \cite{Hallatschek2023, Tang2024}. 
These processes are exemplary cases where 
the motion of individuals is self-propelled \cite{Romanczuk2012, tenHagen2011, Fily2012, Digregorio2018, caprini2020}, 
making such biological phenomena key applications 
of the emerging framework of proliferating
active matter \cite{Hallatschek2023, Giomi2013, Mishra2017, Kalziqi2018}.

Inspired by these biological processes, 
we recently introduced a simple model of moving, proliferating finite-size particles \cite{Almodovar2022, Almodovar2024}. In this model, particles interact through a pairwise potential, and we implemented a mechanism of space exclusion during birth processes. Specifically, reproduction events cannot proceed if there is not sufficient space around the individual selected for reproduction. While particle size is relevant for these processes, particles of the same size were otherwise indistinguishable from one another. That is to say, the only particle feature that matters in this earlier model is the size of the particle. We explored both passive and self-propelled motion, analyzing the spatial structure of a single population of disks \cite{Almodovar2022, Digregorio2018, Kapfer}, and investigating the conditions for coexistence in a binary mixture of two disk types with different sizes \cite{Almodovar2024, Biben1991, Frenkel1992}. Here, we generalise this model to include particles of different types or species. To model the interaction between different types of particles we make use of ideas of game theory.

Game theory has been widely applied to model social and biological systems, where an individual’s success depends on interactions with other individuals \cite{Smith1982, Hofbauer1998, Nowak2006, Szabo2007, ROCA2009}. In a social setting each individual plays a certain strategy at any one time, and then receives a reward from interactions with other individuals. The reward depends on the strategy the individual played and on those played by the interaction partners. Individuals can then update their strategy based on success. In the context of biology, individuals in evolutionary games carry a fixed type representing e.g. a given species or genotype. These then interact with other members of the population, and biological fitness (i.e., the propensity to reproduce) is determined from these interactions. 

Much of the existing work on evolutionary games assumes either well-mixed populations or stylised interaction graphs. In biological processes such as tissue formation or biofilm growth, particles are mobile, and dynamic interaction graphs are determined by this motion. This realization has driven the development of proliferating active matter, integrating finite-size, active motion, and birth-death dynamics \cite{Marchetti2013, Romanczuk2012, Hallatschek2023}. Game theoretic interaction between mobile particles has been studied in \cite{pigolotti2013,Mayer2021,Yu2010,FREY2010,herrerias2019}.

In a model with game theoretic interaction between particles, the reproduction rate of a particle is influenced by the number of particles of both its own type and the other type in its surroundings \cite{Smith1982, Nowak2006, Sigmund1993}. Such interactions are inspired by ecological and evolutionary processes, where competition and cooperation are main drivers of population dynamics \cite{Hofbauer1998, Levin1974, Doebeli2004}. These interactions can lead to different outcomes: coexistence of the two populations, bistability, and dominance of one of the particle types or species \cite{Hofbauer1981, Cressman2003}. In this work, we will analyze how the parameters such as diffusivity, activity, or growth rates, affect the spatial distribution of particles, extinction times, and the response of the system to introducing a new particle \cite{Hallatschek2007, Melbinger2010, Szabo2007} in the different scenarios. Our approach highlights how active motion and spatial constraints can reshape game dynamics, offering insights which could be useful for phenomena ranging from tissue formation to microbial colony expansion.

The structure of the paper is the following. In
Sec. \ref{Sec:model}, we present  the model of mobile disks undergoing
birth and death subject to game-dynamical interaction. 
The main outcomes are presented
in Sec. \ref{Sec:results}. We then summarise and discuss our findings in Sec. \ref{Sec:summary}.

\section{Model and numerical algorithm}
\label{Sec:model}

The model is similar to that in \cite{Almodovar2022}
but with two types of particles 
(or species), which we will call type A 
and type B. The system describes a two-dimensional space with $N(t)=N_A (t) + N_B (t)$ interacting disks, all with diameter $\sigma$. 
We consider the overdamped 
limit and take the friction
coefficient to be equal to unity 
for both species. 
The motion of the disks is  as follows ($\br_i$ is the location of particle $i$),
\begin{equation}
\dot{\br_i}= {\bF}_i + {\bF}^{act}_i + \sqrt{2D_i} \boldsymbol{\zeta}_i (t), \ \ i=1,...,N(t).
\end{equation}
If disk $i$ is of type A then $D_i = D_A$, and if it is of type B then $D_i = D_B$. The variables
$\{\boldsymbol{\zeta}_i\}$ are independent Gaussian noise
vectors satisfying 
$\langle \boldsymbol{\zeta}_i \rangle= 0$, 
$\langle {\zeta}_{i,k} (t) { \zeta}_{j,l} (t') \rangle = \delta_{i j} \delta_{k l}\delta (t-t')$ 
($k$ and $l$ are the entries  of the two-component vectors
$\boldsymbol{\zeta}_i$ and $\boldsymbol{\zeta}_j$). 
The diffusivities $D_A$ and $D_B$ are taken
as parameters of the model.

The finite size of the disks is simulated using
a truncated Lennard-Jones potential  
so that the force on particle $i$ 
resulting from the 
interaction with 
the rest of particles (of any type)
is $\bF_{i} = -{\bf \nabla}_i \sum_{j\neq i}  U(|\br_i -\br_j|)$,
where the potential is given by (with 
$r=|\br_i -\br_j|$)
\begin{equation}
U(r) = 4 \varepsilon \Bigl[\Bigl(\frac{\sigma}{r}\Bigr)^{12}
- \Bigl(\frac{\sigma}{r}\Bigr)^6\Bigr] + \varepsilon,
\end{equation} 
if $r< 2^{1/6} \sigma$, 
and $U(r)=0$ if $r > 2^{1/6} \sigma$. The parameter
$\varepsilon$ is an energy scale \cite{LB_rule_Book1,LB_rule_Book2}.

Particles can be self-propelled, 
which is  
 modeled by using active forces 
\begin{equation}
\bF^{act}_i = v \,\bn[\theta_i (t)]
\label{eq:forceact}
\end{equation} 
of constant modulus $v$ 
(typically called activity or velocity),
 and with a direction given by the unit vector
$\bn (\theta_i) = ( \cos \theta_i, \sin \theta_i )$.
The angle $\theta_i$ for disk $i$ performs diffusive motion,
$\dot{\theta}_i (t) = \sqrt{2 D_r} \eta_i (t)$.
The term $\eta_i$ represents a zero-mean Gaussian noise with
$\langle \eta_i (t) \eta_j (t')\rangle = \delta_{ij} \delta (t - t')$,
and $D_r$ is the rotational diffusion coefficient.

Disks may randomly self-replicate 
or die, so that
the number of disks of each 
type, $N_A (t)$ and $N_B (t)$,
can change with time. 
The birth and death dynamics are subject to game dynamical interaction and occur as follows (see also \cite{Almodovar2022} for a similar model without game dynamics):
\begin{itemize}
\item [(i)] Each existing particle 
dies with constant per capita rate $\delta$.
Particles that die
are removed from the system.
 
\item[(ii)] A potential
 reproduction of particle $i$ 
 is triggered with rate
\begin{equation}
    \beta_i^{X}=\beta_0 [1+ \alpha \tanh( \Pi_i^{X})],
    \label{birth_rate}
\end{equation}
where $\Pi_i^X$ is the payoff to particle $i$ at the time of the proposed event. The superscript $X\in\{A,B\}$ indicates that this payoff depends on the type of the particle. As explained below $\Pi_i^{X}$ also depends on the number of types of particles in the vicinity of the particle proposed for reproduction. The parameter $\beta_0$ in Eq.~(\ref{birth_rate}) describes a baseline reproduction rate, in absence of other particles. This will become clear below when we define the $\Pi_i^X$. The coefficient $\alpha$ is an `intensity of selection', and modulates how strongly the reproduction rate $\beta_i^X$ depends on the payoff to particle $i$. For $\alpha=0$ selection is neutral, i.e., there is no dependence on payoff. 

\item[(iii)] In a reproduction event the parent particle produces one offspring. The offspring is of the same type as the parent, i.e., we do not consider mutation. Crucially, proposed birth events are only executed if sufficient space exists around the parent particle to accommodate the offspring without overlapping with other disks.  If there is no space, no birth event occurs. This means that not all potential reproduction events complete. We implement this as in \cite{Almodovar2022, Almodovar2024}.
\end{itemize}

It remains to define the payoffs $\Pi_i^X$. These are
\begin{equation}
\begin{aligned}
    \Pi_i^A=N_i^A a + N_i^B b,\\
    \Pi_i^B=N_i^A c + N_i^B d,
\label{matrix_generica}
\end{aligned}
\end{equation}
where $N_i^A$ is the number of particles of type A within a fixed radius $R$  around particle $i$, and $N_i^B$ is the number of type-B particles within this radius. The coefficients $a, b, c$ and $d$ in Eq.~(\ref{matrix_generica}) are model parameters, they specify the payoff matrix of the underlying game \cite{Hofbauer2003}. The setup reflects the localized nature of interactions observed in many biological and ecological systems, where the success of an individual is influenced by the immediate environment \cite{Nowak2006, Levin1974}. For instance, microbial competition in biofilms or the dynamics of resource allocation in spatially constrained populations are strongly influenced by local densities and proximity to competitors or cooperators \cite{Doebeli2004, Szabo2007}. 

In the model, the birth rate $\beta_i^X$ will change in time,  as the set of particles around a focal particle will change as the particles move. We employ a Gillespie-type thinning algorithm to simulate the population dynamics \cite{Lewis1979,Ogata1981}. We always assume that 
the baseline birth rate, $\beta_0$, is 
larger than the death rate 
(both are taken 
to be equal for all the disks), 
 $\beta_0 >\delta$. 
$\alpha$ is a positive parameter
between $[0,1]$ to ensure that the birth rate is always positive.  
Changing its value does not qualitatively alter our results, as confirmed by simulations not shown in the paper. Therefore, we set $\alpha=1$ throughout.

The parameters that determine 
the interaction, $a,b,c,d$,
can be positive  
or negative,  
corresponding to well-known evolutionary dynamics involving 
two-species competition. We can distinguish between the following scenarios 
\cite{May1977, Nowak2004}:
\begin{enumerate}
\item {\it Coexistence.} Populations of both types A and B can coexist in stable proportions if $a\leq  c$ and $d \leq b$. In this situation a population entirely consisting of type-A particles can be invaded by type B, and vice versa.

To simplify the analysis, we consider a symmetric scenario $a=d=-b=-c$ and $a<0$.  Such symmetry in the interaction matrix is commonly used in game theory to reduce the complexity of parameter space while retaining the essential dynamics of competition and coexistence \cite{Smith1982, Nowak2006, Hofbauer1998}. Under this assumption, the payoffs for each particle type can be written as:
\begin{equation}
\begin{aligned}
\Pi_i^A=a (N_i^A - N_i^B) =a \Delta^{AB}_{i}, \\
\Pi_i^B=a(N_i^B - N_i^A) = a \Delta^{BA}_{i}.
\label{matrix_a}
\end{aligned}
\end{equation}
The payoffs of each particle type thus depend on the difference in the number of particles of each type. We have $\Delta^{AB}_i=-\Delta^{BA}_i$. Since for coexistence $a$ is negative, particles tend to mix more with each other rather than segregating into separate regions of space. As a result, this leads to spatially homogeneous configurations, with no segregation of the two types. Both particle types are evenly distributed throughout the system.

\item {\it Bistability.} Each strategy is the best response to itself. Depending on initial conditions, either type $A$ or type $B$ will eventually vanish. 

To study this scenario we set again $a=d=-b=-c$, but now with  $a>0$. The payoffs can again be written as in Eq. (\ref{matrix_a}),
but $a$ is positive. Particles tend to cluster with others of the same type, potentially leading to spatial segregation. 

\item  {\it Dominance.} We focus on the dominance of type B (but by symmetry the analysis can easily be adapted to dominance of type A). We assume that the payoff to $B$ is always higher than that to $A$, given a set of neighbours, i.e., we have  $c>a$ and $d>b$. In this case, type-$B$ particles dominate the 
system, and type-$A$ particles become extinct eventually.  Unlike in the case of bistability this occurs independently of the initial mixture of $A$ and $B$ particles. 

\item {\it Neutrality.} This corresponds to $a=b=c=d$ (neutrality can alternatively also be achieved by setting $\alpha=0$). Both 
types are identical in every aspect which
means so that we have only one 
type of particle. We will not investigate this further in this paper.
\end{enumerate}

For each of the first three scenarios, we perform different studies
varying the different parameters of the model, such as the diffusivities, activity, or the birth parameters.
In the case of coexistence we will focus on
the steady spatial distribution,
for the other scenarios we will mainly
analyse the time until one species goes extinct, and the fixation or survival probability
for particles of a given type.

\section{Results}
\label{Sec:results}

We consider a two-dimensional box of length $L_s=70$ with
periodic boundary conditions, $\delta=0.01$, and
$\sigma=1.0$. We set $\varepsilon=1.0$ and $D_r = 1.0$. These choices do not affect the qualitative behavior of the results. Also, we fix the activity parameter at $v = 1$ when we do not study the
dependence on $v$.
 Simulation  results are independent of $\delta$
whose role is mostly to set the time
scale needed to reach the steady state 
\cite{Almodovar2022}.
Unless otherwise stated,
we start with $50$ particles of each type,
together occupying around 
$2\%$ of the total area.
The packing fraction 
for particles of either type is
given by
$\phi_{X}(t)=N_{X}(t)\pi (\sigma/2)^2/L_s^2$,
where $N_X(t)$ is the number of 
particles of type $X\in\{A, B\}$ at time $t$.
The radius R, used to determine what other particles a given member of the population can interact with in the game, is $R=2\sigma$. This radius determines $N_i^A$ and $N_i^B$ in Eq.~(\ref{matrix_generica}). Choosing a radius that is too large would effectively remove the notion of space (as all particles would interact with all other particles in the game). If $R$ is too small, no game dynamical interaction occurs.

We study the coexistence in Sec.~\ref{sub:coex}, 
the bistable scenario in Sec.~\ref{sub:bist}, and the situation in which type-$A$ dominates the game in Sec.~\ref{sub:Domination}.

\subsection{Coexistence}
\label{sub:coex}

\begin{figure*}[hbt!]
\begin{center}
\includegraphics[width=0.7\linewidth]{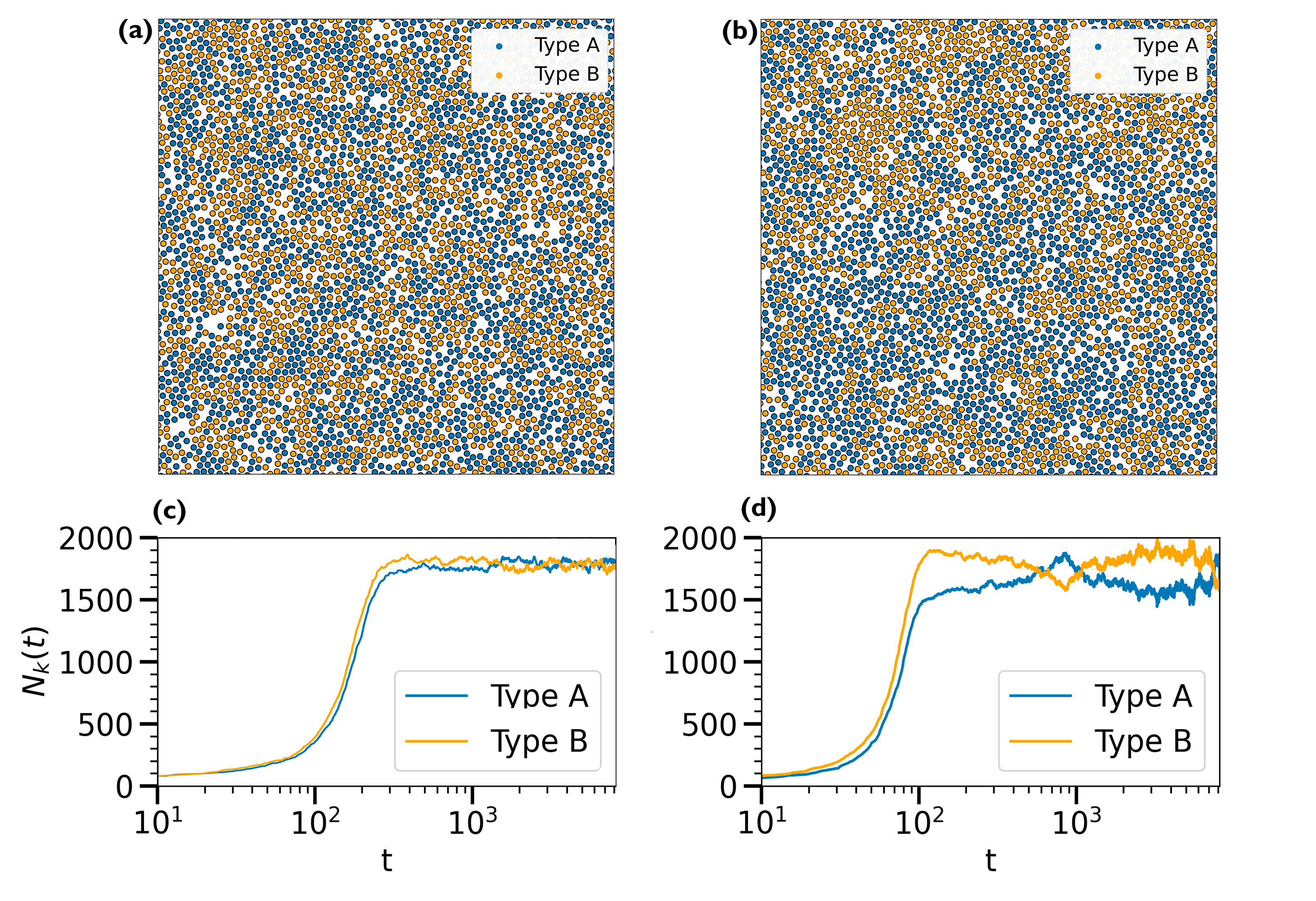}
\caption{
Panels (a) and (b) show snapshots of the long-time spatial 
distribution of active disks in space.
Panels (c) and (d) show the corresponding time evolution 
of the packing fraction starting 
with a random configuration of 50 disks of each type. 
The snapshots in
(a),(b) are taken at the end of the time series in (c),(d), respectively. We have used $a=-1$ for (a) and (c), and $a=0$ for (b)-(d). The remaining parameters are  $\beta_0=0.05,\delta=0.01,D=0.05,v=1.0$.}
\label{Fig:Snapshot_coex}
\end{center}
\end{figure*}
\subsubsection{Setup and general behaviour}
We start with the scenario $a = d = -b = -c$ and $a < 0$. Using Eqs.(\ref{birth_rate}) and (\ref{matrix_a}), the reproduction rate of a given particle increases when it is surrounded by particles of the other type, enhancing mixing. This is observed in Fig.~\ref{Fig:Snapshot_coex}(a) (for $a = -1$), compared to Fig.~\ref{Fig:Snapshot_coex}(b), where we show the case $a=b=c=d=0$, i.e., a system with no game dynamcal interaction. We will discuss this in more detail in Sec.~\ref{sub:pd} below.

 In Fig. \ref{Fig:Snapshot_coex}~(c) and (d) we plot the time evolution of the packing fractions. Both particle types coexist. For $a = -1$ [Fig.~\ref{Fig:Snapshot_coex}(c)] fluctuations in time are small, and the populations remain nearly balanced. In contrast, for $a = b=c=d=0$. Fig.~\ref{Fig:Snapshot_coex}(d) shows that the populations of both types fluctuate around an approximately equal balance, with moments in time where one type temporarily exceeds the other. These fluctuations do not indicate a permanent imbalance but highlight the dynamic nature of the system in the absence of game dynamical interactions.

We now study in more detail the 
spatial distribution of disks, 
and the response of the steady system
when a new particle is added. Given that coexistence is the ultimate outcome, we cannot measure any fixation time or probability (i.e., the time it takes for the descendants of a new particle to dominate the system, or the probability with which this happens). However, we can obtain the survival probability for the lineage of a particle invading a population consisting entirely of the other type.

\subsubsection{Mixing and local neighborhood composition}
\label{sub:pd}

\begin{figure}[hbt!]
\begin{center}
\includegraphics[width=1.0\linewidth]{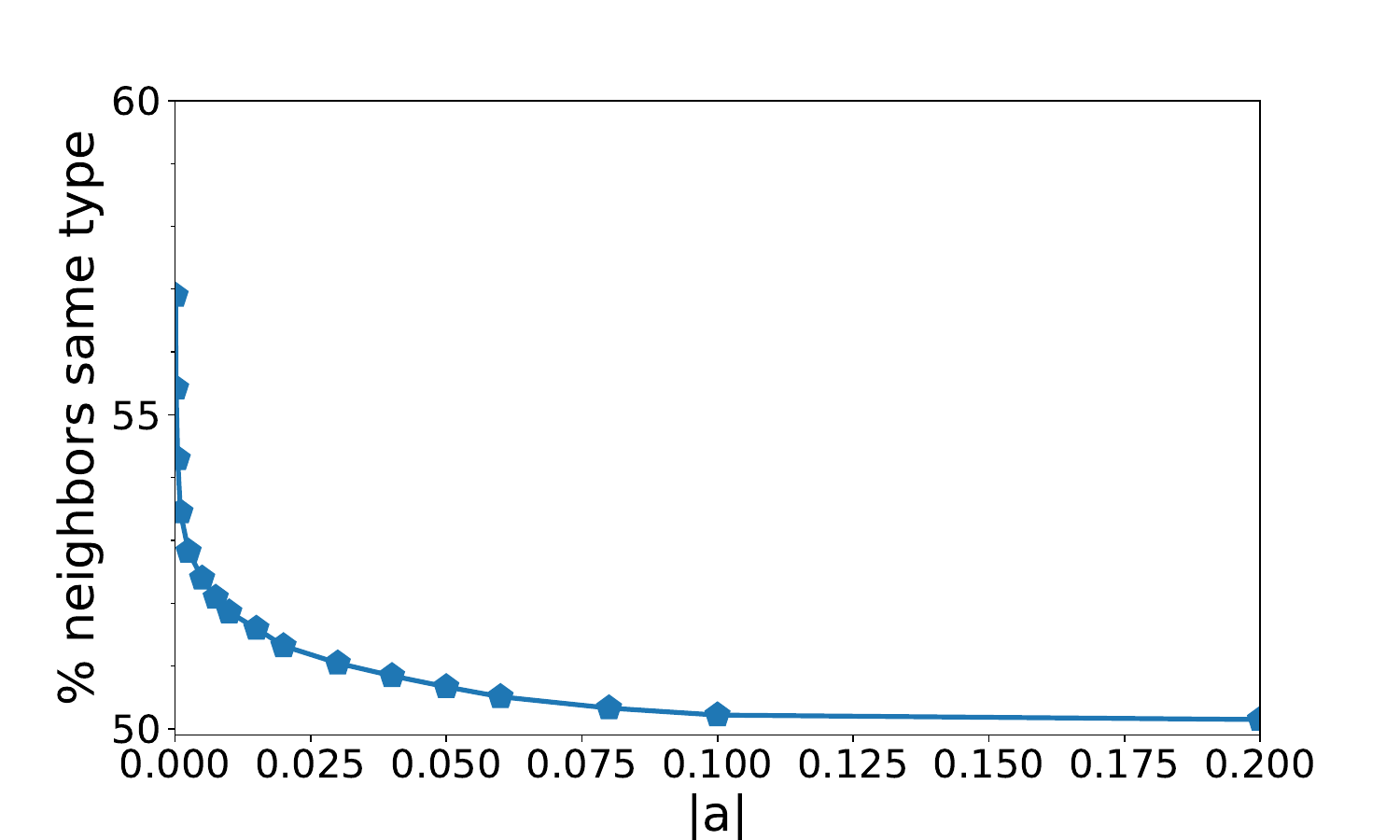}
\caption{Percentage of neighbors of the same type vs $a$. 
The remaining parameters are  
$\beta_0=0.05,\delta=0.01,D=0.05,v= 1$. 
This is computed by averaging 
the percentages over all particles
in the system, considering a 
time window under steady-state conditions
and multiple realizations.
}
\label{Fig:perc_nn}
\end{center}
\end{figure}

In Fig.~\ref{Fig:perc_nn}, we plot the percentage of particles of the same type surrounding a given particle within a radius $R = 2\sigma$. An increasing value of the absolute value of $|a|$ indicates stronger interactions between particles. These interactions are here of the coexistence type, and with stronger interactions, the system becomes more mixed. This is reflected in a decrease in the percentage of same-type neighbors, showing that particles are increasingly surrounded by particles of the other type.

The most mixed situation would correspond to a fraction of $50\%$ on the vertical axis of Fig.~\ref{Fig:perc_nn}, indicating that a particle is equally likely to be surrounded by particles of its own type or of the other type. We find that the percentage of same-type neighbors never falls below 50\%. This is because reproduction events place the offspring near their parent particles and because offspring and parent are of the same type. Thus, while the system becomes increasingly mixed with stronger interactions, there will still be a tendency for a particle to be surrounded by particles of its own type.

\begin{figure}[hbt!]
\begin{center}
\includegraphics[width=1.0\linewidth]{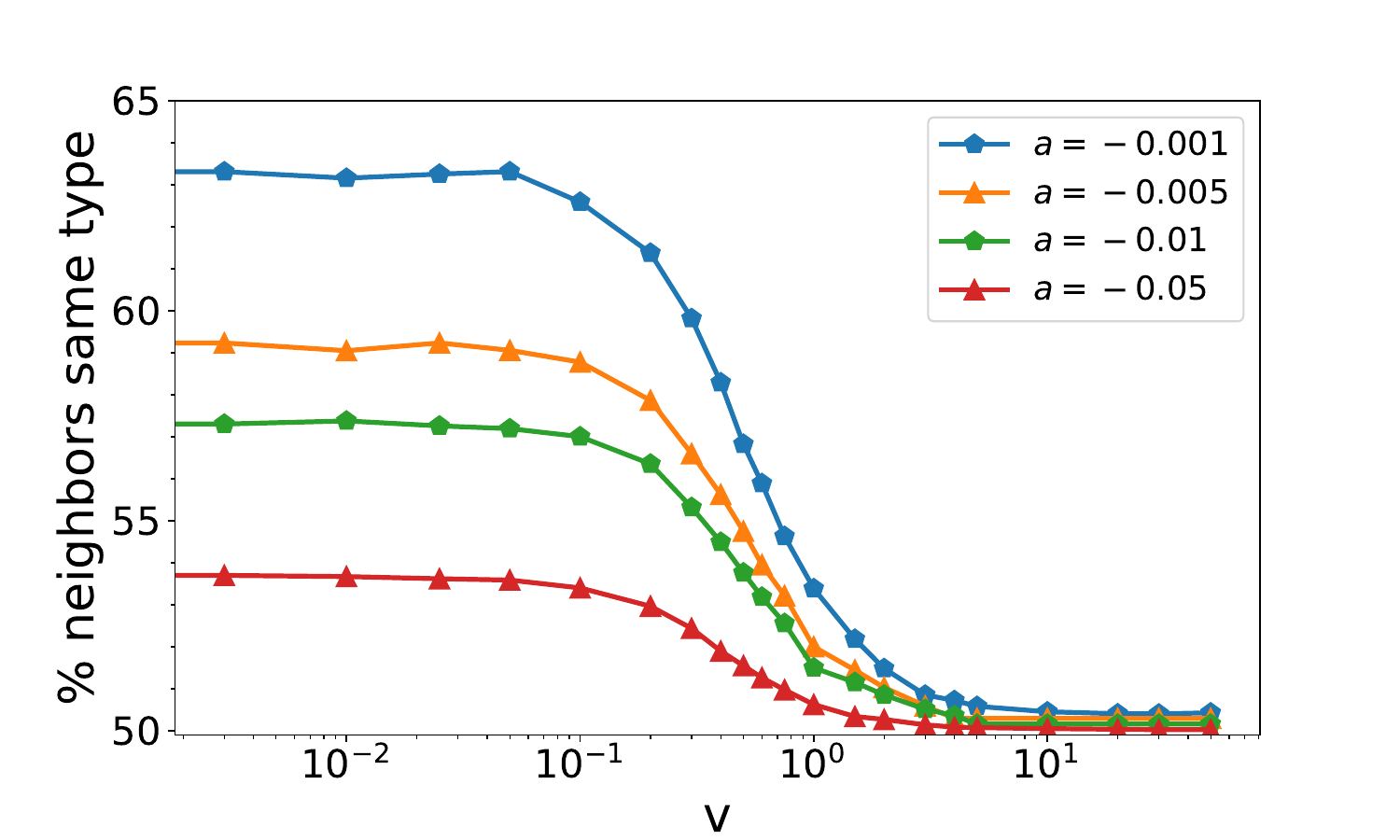}
\caption{Percentage of neighbors of the same type vs $v$ 
for different values of $a<0$.
The remaining parameters are  
$\beta_0 = 0.05, \delta=0.01, D=0.05$. }
\label{Fig:perc_nn_act}
\end{center}
\end{figure}

In Fig.~\ref{Fig:perc_nn_act}, we illustrate how mixing depends on the self-propulsion velocity $v$, again by plotting the percentage of neighbors of the same type for different fixed values of $a<0$. At low propulsion velocities, particles move slowly relative to the rate of reproduction, leading to limited mixing and a relatively high proportion of same-type neighbors. As $v$ is increased, particles explore larger areas and mix more effectively, with the percentage of same-type neighbors approaching 50\% at high velocities. This behavior is consistent with the effectively enhanced diffusion coefficient associated with increased self-propulsion velocity \cite{Caprini_Diff}.

\begin{figure}[hbt!]
\begin{center}
\includegraphics[width=0.8\linewidth]{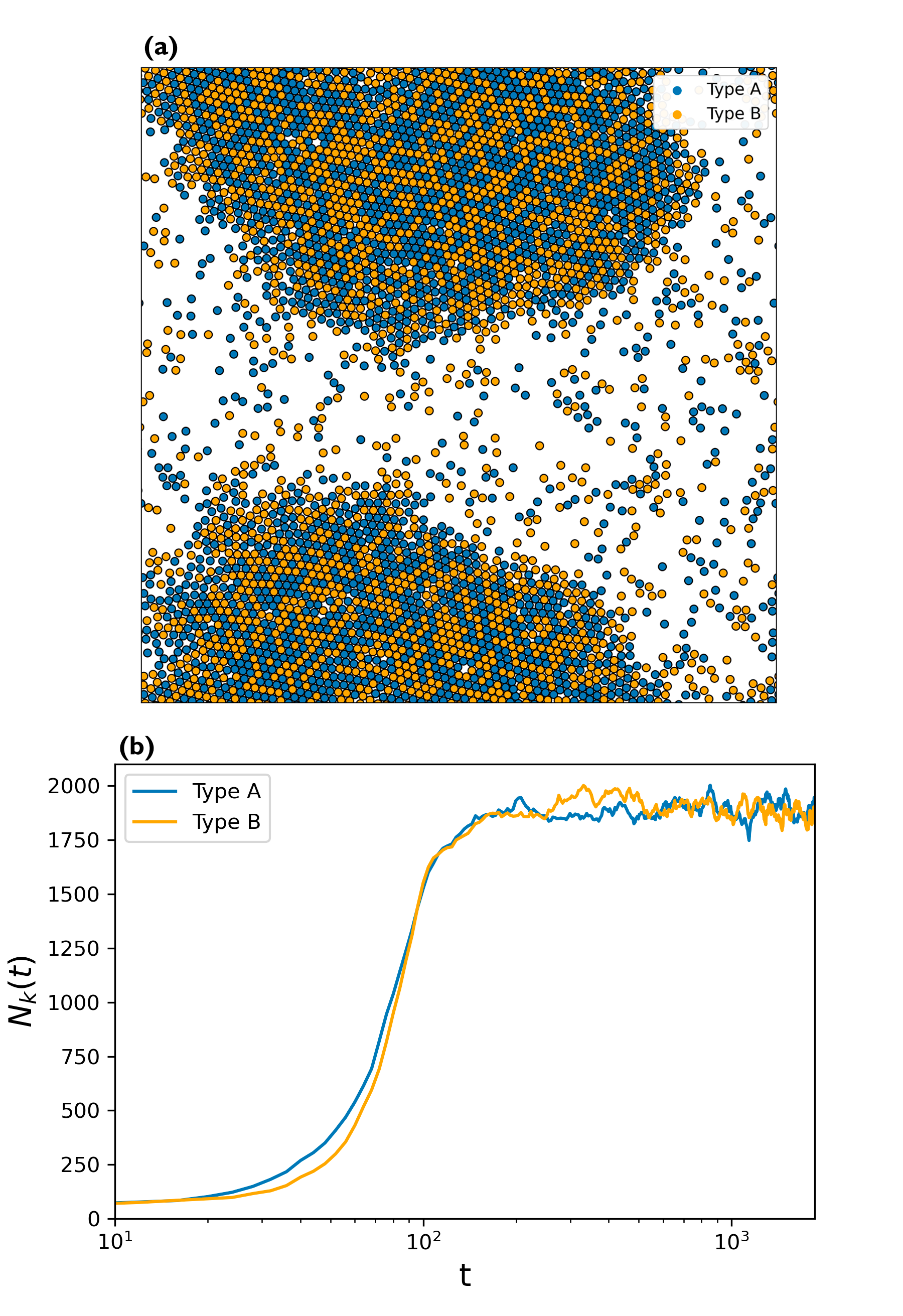}
\caption{(a) Snapshots of the long-time spatial 
distribution of active disks in space for $a=-1$.
(b) The corresponding  temporal evolution 
of the packing fraction starting 
with a random configuration of 50 disks of each type. 
The snapshots in
(a) is taken at the end of the time series in (b).
The remaining parameters are  $\beta_0 = 0.05,\delta=0.01,D=0.05,v= 50 $.}
\label{Fig:Snapshot_coex_MIPS}
\end{center}
\end{figure}

We show a typical configuration for large velocities ($v=50$) and strong interactions of the coexistence type ($a=-1$) in Fig.~\ref{Fig:Snapshot_coex_MIPS}. 
The system shows motility-induced phase separation (MIPS), with dilute and densely packed areas of the system. A degree of mixing is seen both in the dilute and dense phases. In Fig. \ref{Fig:Snapshot_coex_MIPS}(b) we show how the numbers of type-$A$ and type-$B$ particles evolve in time. As seen, the system contains approximately equal numbers of particles of either type at all times.

\subsubsection{Factors affecting coexistence probability}
\label{sub:fp_coex}

\begin{figure}[hbt!]
\begin{center}
\includegraphics[width=1.0\linewidth]{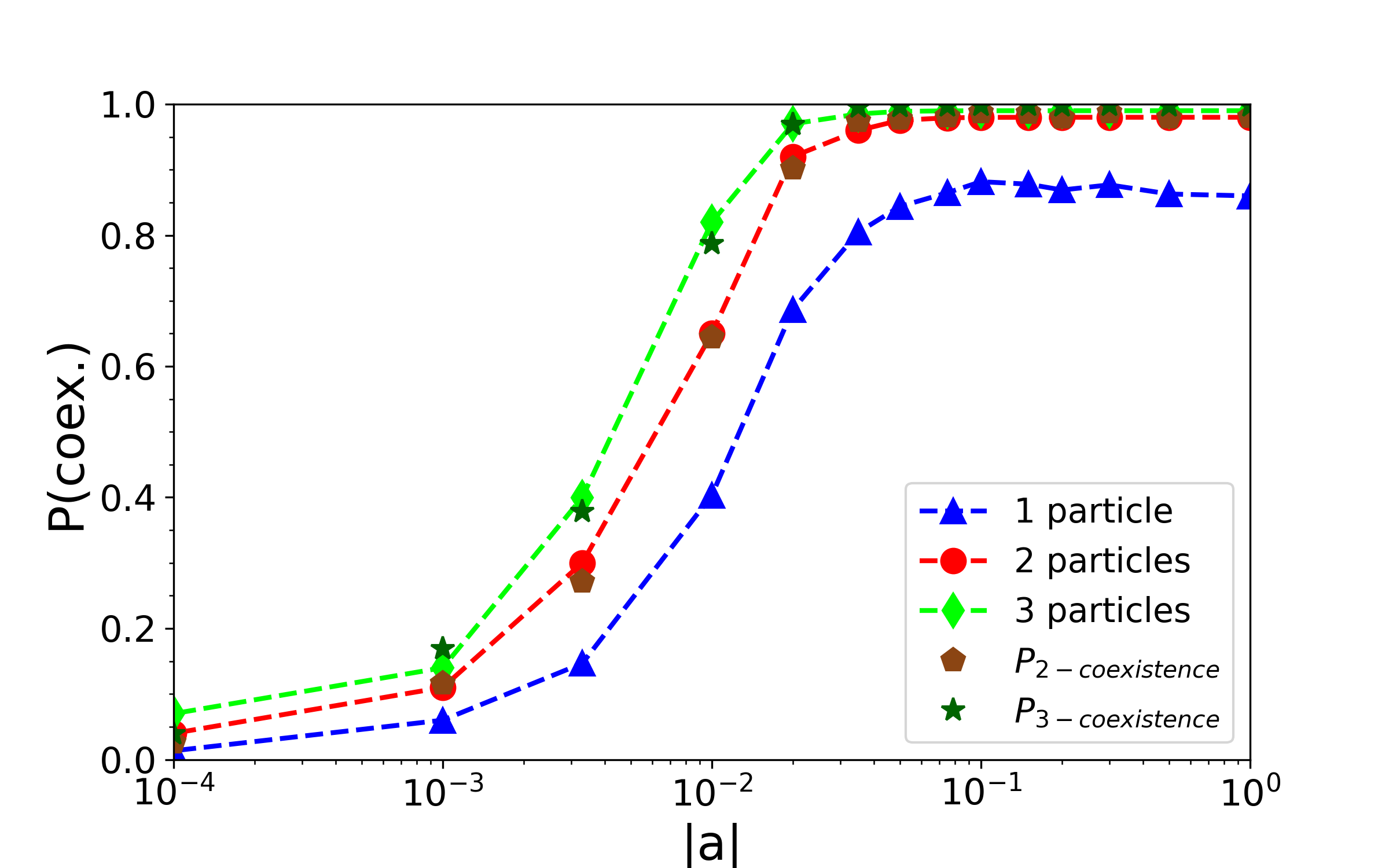}
\caption{Probability of reaching
coexistence vs $|a|$. 
It is computed for a  
different number of initial particles,
i.e., if we start with  $1$, $2$ or $3$ particles. 
Brown pentagons and green stars are obtained 
using Eq.~(\ref{Eq:P_k}) from the data 
of only introducing  one new particle. The remaining parameters are  $\beta_0 = 0.05,\delta=0.01,D=0.05,v=1.0$.
The percentage is computed by averaging from 300 simulations for each value of $|a|$.
}
\label{Fig:prob_coex}
\end{center}
\end{figure}

In this subsection, we report results from numerical
experiments involving the introduction of foreign disks. 
We begin with a system containing only type-A particles, and introduce a few type-B particles  at  random positions. Given that $a$ is negative, 
particles surrounded predominantly by the opposite type experience an increase in their birth rate. We thus expect the newly introduced type-B particle to have a high chance of reproducing, along with its offspring. We can then ask under what conditions the lineage of the newly introduced particle type survives in the long run. Results are shown in Fig.~\ref{Fig:prob_coex}.

For very small values of $|a|$, the probability of achieving coexistence is negligible, as shown in Fig.~\ref{Fig:prob_coex}. This is true regardless of whether the initial number of foreign particles is $1$, $2$, or $3$. However, as $a$ becomes more negative ($|a|$ becomes larger), the probability of coexistence increases, and the impact of the initial number of foreign particles becomes more pronounced. When a single type-B particle is introduced, the maximum coexistence probability, even for very negative $a$, reaches about $85\%$. This indicates that invasion is not successful with probability one, even if the drive by the game dynamics towards coexistence is strong. With only a single type-B particle initially, there is a chance that this particle may die before reproducing, leaving no particles of type B in the system. By introducing three type-B particles, the coexistence probability increases to values that are indistinguishable from $100\%$ in our simulations. The risk of complete extinction is then negligible when the effect of the game is strong (large $|a|$).

We now study this in more detail. Starting from a small number of type-B particles, there are two possible outcomes, either both types of particles 
coexist in the long run, or  the newly introduced type of particle goes extinct. If there are $k$ particles of type B initially we write $P_{\rm k-coex}$ for the probability of the former outcome, and $P_{\rm k-ext}$ for the latter. These two probabilities sum to one. We now further assume that the extinction of each of the $k$ B-particles occurs independently from the other B-particles. This means to assert that $P_{\rm k-ext}=(P_{\rm 1-ext})^k$. From this we find
\begin{equation}
    P_{\rm k-coex} = 1- (P_{\rm 1-ext})^k.
    \label{Eq:P_k}
\end{equation}

Thus, the probability of
 coexistence after introducing $k$
type-B particles can be predicted 
based on the outcome of introducing
just a single type-B particle. This 
is confirmed in Fig.  \ref{Fig:prob_coex}.

\begin{figure}[hbt!]
\begin{center}
\includegraphics[width=1.0\linewidth]{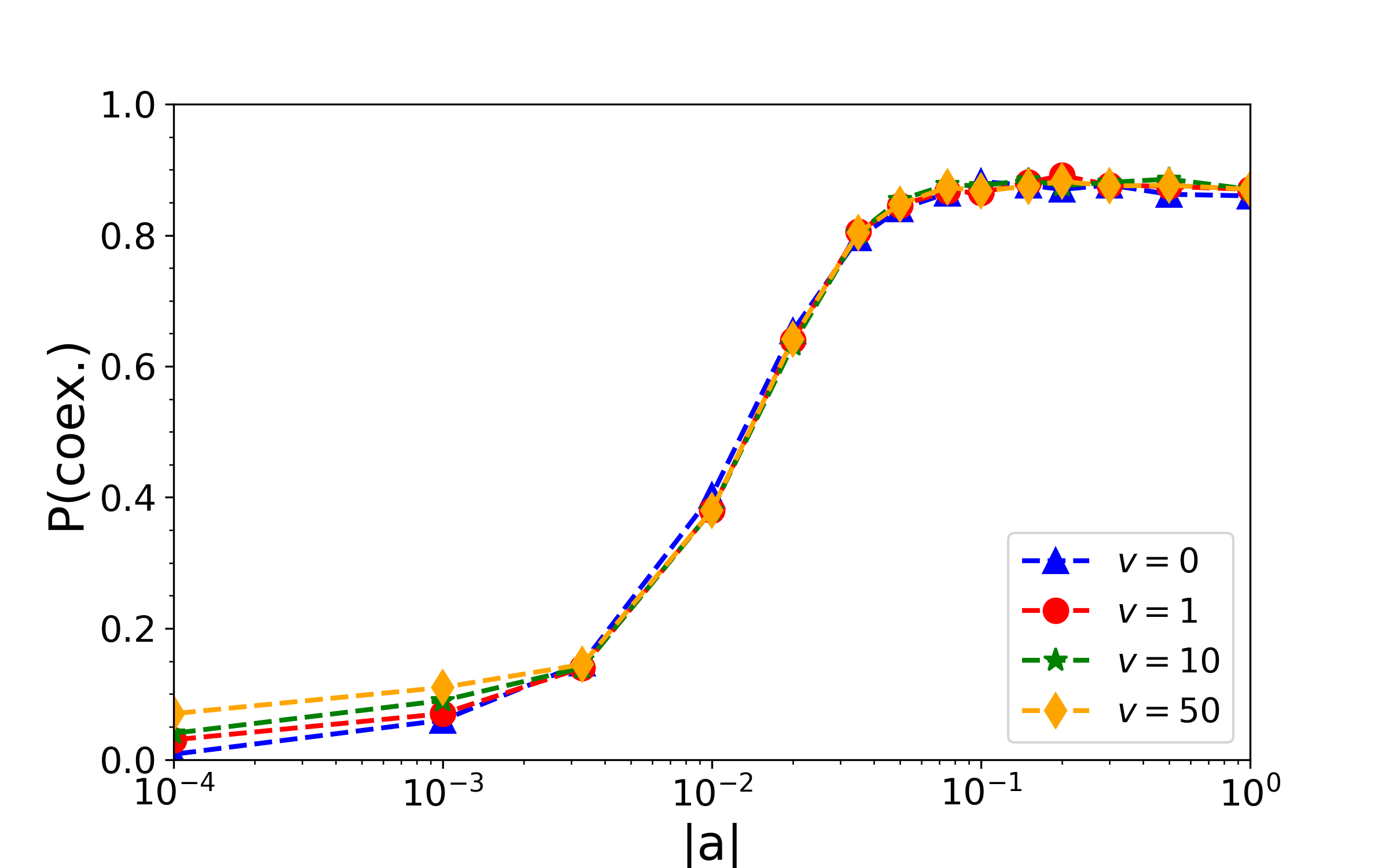}
\caption{Probability of reaching
coexistence vs $|a|$ and different 
values of the activity $v$.  
The remaining parameters are 
$\beta_0=0.05,\delta=0.01,D=0.05$.}
\label{Fig:prob_coex_act}
\end{center}
\end{figure}

The influence of the activity on the probability of coexistence is shown in Fig.~\ref{Fig:prob_coex_act}. Here, we always start with a single foreign particle. For large values of $|a|$ (i.e., strong pull towards coexistence), activity has almost no effect, and an $85\%$ survival probability is reached for the foreign type of particles, independently of $v$. For small values of $|a|$ (i.e., less selection pressure), the influence of activity is small but not negligible. In fact, larger activity increases the probability of coexistence because it enhances the mixing of particles in space, i.e., there is an increased tendency for the invaders to be surrounded by particles of the opposite type. As a consequence, the invading particles have a higher probability of reproducing. Increased activity also allows particles to explore larger areas and encounter regions populated by the opposite type, thereby promoting reproduction and increasing the probability of coexistence.

Due to the symmetry of the chosen values  for $a,b,c,d$, we have found that coexistence  corresponds to approximately $50 \%$ of the  total population for each type.  However, with a different choice of these parameters, 
this balance may shift, and  different percentages of coexistence 
for the two types can emerge for long times.

\subsection{Bistability}
\label{sub:bist}
\subsubsection{Setup and general behaviour}
We now study the case when $a=d=-b=-c$ and $a>0$. 
For positive values of $a$, disks tend to cluster with others of their own type.  This occurs because  particles surrounded by the same 
type are more likely to reproduce, as  indicated by 
Eqs.~(\ref{birth_rate}) and (\ref{matrix_a}). 
If a particle finds itself in the middle
of a cluster of the opposite type, it is more likely to die before it can reproduce. In the long-time there is no coexistence and
only one type of particle survives (which type this is, depends on the initial conditions).

\begin{figure*}[hbt!]
\begin{center}
\includegraphics[width=0.9\linewidth]{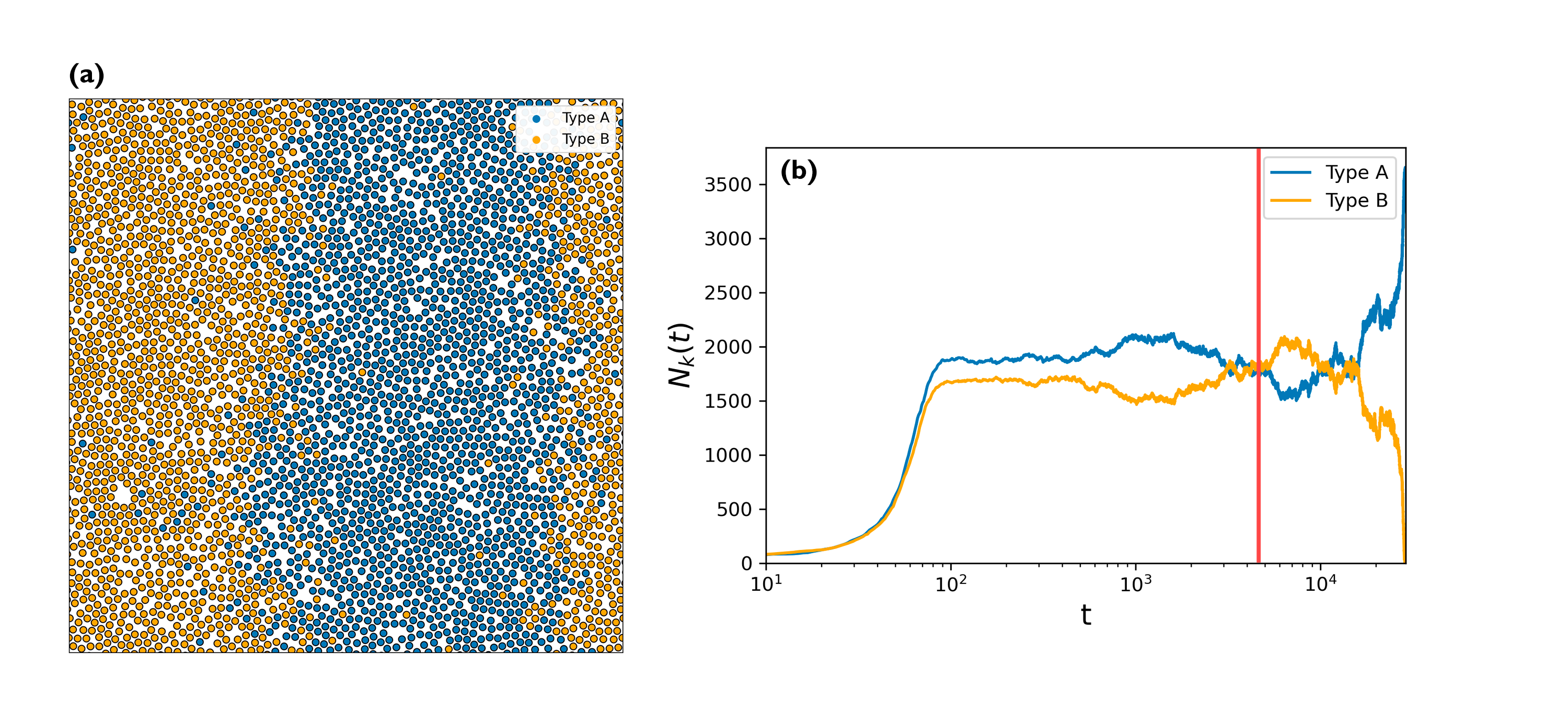}
\caption{
(a) Snapshots of the long-time spatial 
distribution of active disks in space for $a=1$.
(b) The corresponding temporal evolution 
of the packing fraction starting 
with a random configuration of 50 disks of each type. 
The snapshot in
(a) is taken at the time indicated by the red line in (b) .
The remaining parameters are  $\beta_0 = 0.05,\delta=0.01,D=0.05,v=1$.}
\label{Fig:Snapshot_Bist}
\end{center}
\end{figure*}

However, it is reasonable to expect particle-type segregation in transient states.  This is indeed what we find in simulations. 
In Fig. \ref{Fig:Snapshot_Bist}(a) we show 
a transient spatial configuration for 
$a=1$ and $v=1$. A segregation pattern
is observed where particles tend to stay
near those of the same type. 
In Fig. \ref{Fig:Snapshot_Bist}(b) we observe
that one type of particle becomes extinct in the long run (type B in this example).

\subsubsection{Extinction time}
\label{sub:ext_time}
We will now report results for time-to-fixation in the bistable system, and its dependence on model parameters. Unlike in the case of coexistence, we do not start from only a small number of invading particles. This is because such a small group is almost certainly going to be eliminated in the bistable scenario, where the drive is towards monomorphic all-A or all-B states. Instead, we start with an initial 
configuration with equal number of type-A and type-B particles. Eventually, 
one type will go extinct, and we can measure the time that elapses until this happens.

\begin{figure}[hbt!]
\begin{center}
\includegraphics[width=1.0\linewidth]{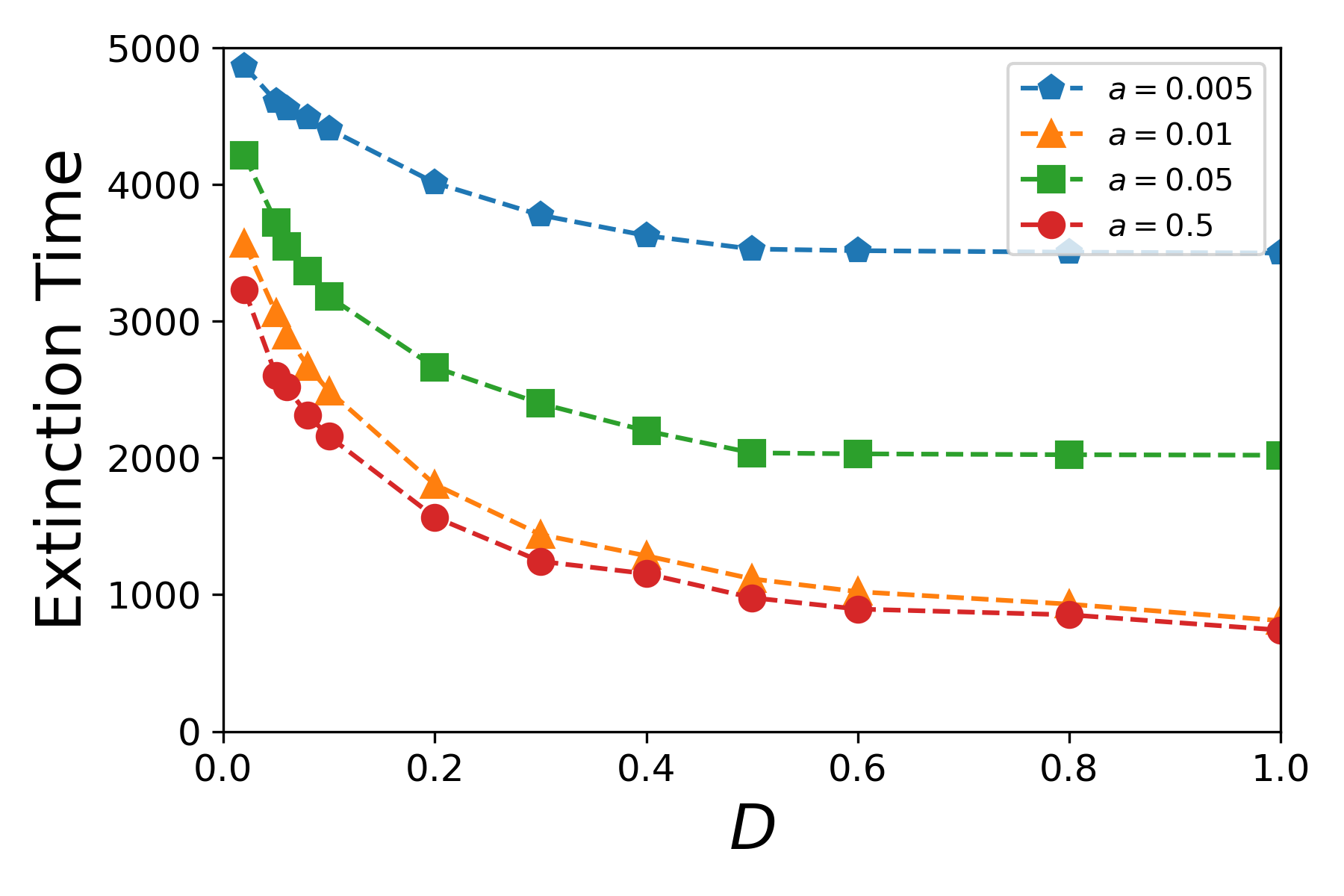}
\caption{Extinction time vs $D$ 
for different values  
$a$. The remaining parameters 
are  $\beta_0 = 0.05,\delta=0.01,v=1$. }
\label{Fig:ext_time_bist}
\end{center}
\end{figure}

In Fig.~\ref{Fig:ext_time_bist}, we plot the extinction time as a function of the diffusivity $D$ for various values of $a$, and $v=1$. As mixing of both types increases due to more movement (higher $D$), the time to extinction shortens. This suggests that larger diffusivity enables one type of particle to dominate the system sooner in the time evolution. Further investigation is needed to fully understand the underlying mechanism.

In all cases, we observe that starting with the system either almost empty or completely full of disks (but with an equal number of each type) results in nearly the same extinction time. This indicates that the outcome is largely independent of the initial condition, as long as the initial number of disks of each type is approximately balanced. This behavior can be explained by the rapid filling of the system with particles (of any type) when the initial number of particles is very small. Once the system is filled, one type goes extinct over a much longer timescale compared to the time required for the system to fill.

If the diffusivities of the two types of particles are different, one might expect that the population with the smaller diffusivity would dominate the system (assuming all other parameters are equal). However,  simulations (not shown here) reveal the opposite: the particle type with smaller diffusivity typically goes extinct. Moreover, in this situation, the extinction time is largely independent of both diffusion values. However, it varies significantly with the parameter $a$. This conclusion is supported by the observation that the type with larger diffusivity can explore the system more effectively and invade areas where the opposite type resides.

\begin{figure}[hbt!]
\begin{center}
\includegraphics[width=1.0\linewidth]{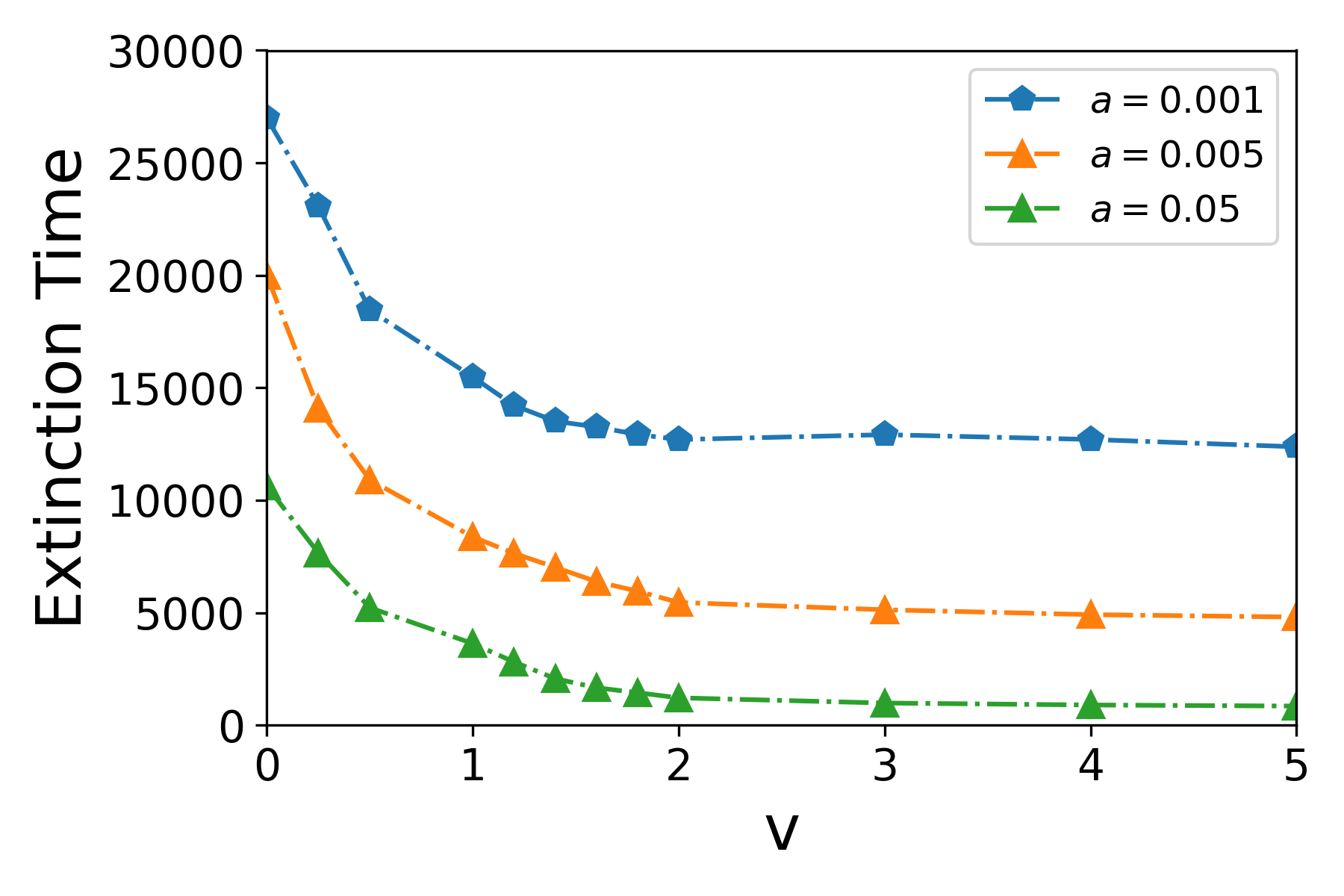}
\caption {Extinction time vs $v$ for different values $a$. The remaining parameters are 
 $\beta_0 = 0.05,\delta=0.01,D=0.05$.}
\label{Fig:ext_time_bist_act}
\end{center}
\end{figure}

We show how the extinction time changes as we vary the activity $v$ in Fig. \ref{Fig:ext_time_bist_act}. Extinction occurs sooner when activity is high. This highlights again that activity and diffusivity have similar roles, at least when $v$ is not too large.

We observe in both Fig. \ref{Fig:ext_time_bist} and
Fig. \ref{Fig:ext_time_bist_act} that the extinction time decreases with increasing parameter $a$. We recall that high values of $a>0$ indicate strong bistability in the game dynamics. The drive towards the monomorphic states is then increased so that extinction occurs faster.

We further note that there is a non-zero minimum extinction time for large $D$ or large $v$ respectively, see the right-hand ends of Figs.~\ref{Fig:ext_time_bist} and
Fig. \ref{Fig:ext_time_bist_act}. The precise value of this plateau depends on $a$, and represents the least 
amount of time required to reach the all-A or all-B states from an initial configuration with equal numbers for both particle types. This time decreases with increasing $a$, again because larger values of $a$ indicate a stronger pull towards extinction by the game dynamics. Conversely, when $a$ is near zero, selection becomes neutral, and thus no particle type has an advantage, independently of the composition of the system.  As a result, any possible extinction occurs purely through random effects (genetic drift), and both types of particle can coexist for longer periods of time than under bistable selection.

\subsection{Domination}
\label{sub:Domination}

\begin{figure*}[hbt!]
\begin{center}
\includegraphics[width=0.9\linewidth]{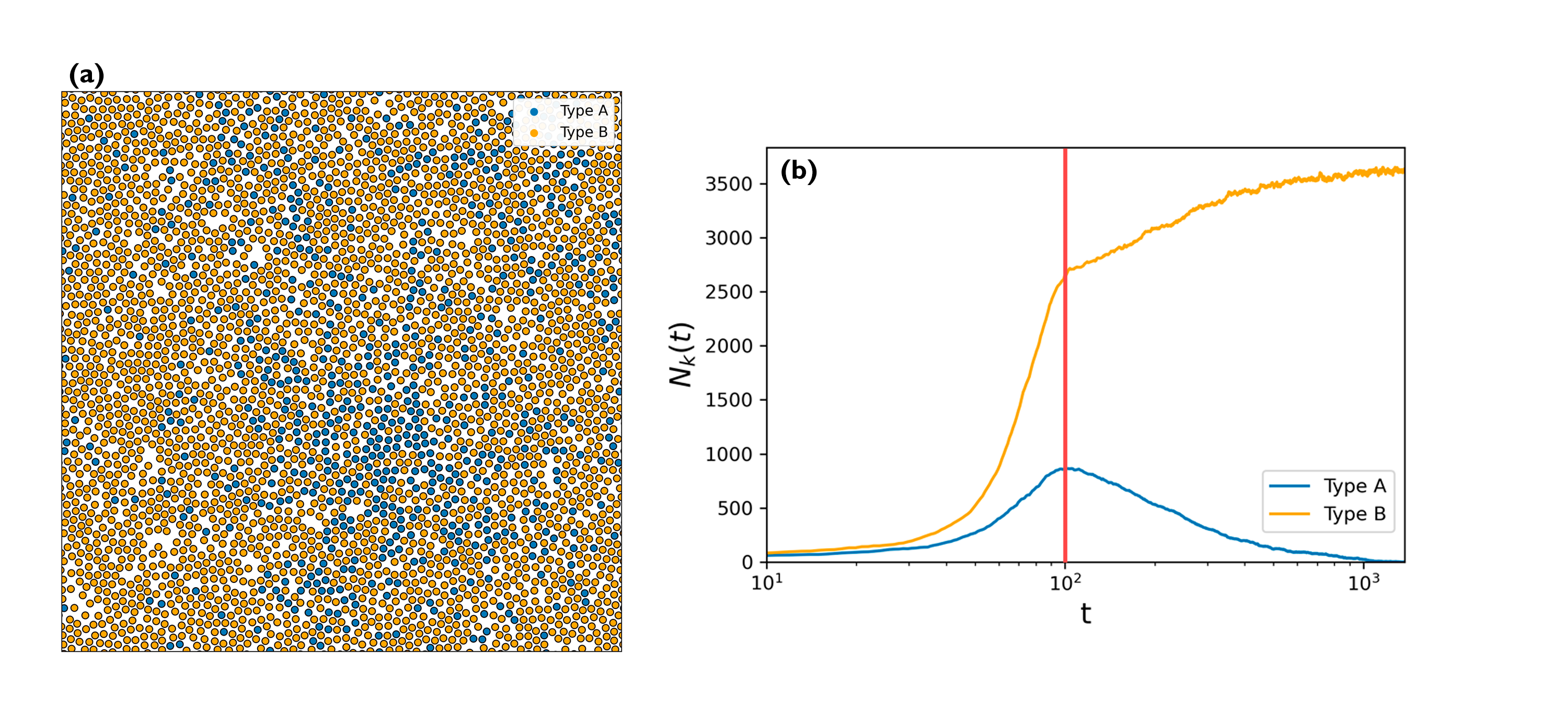}
\caption{
(a) Snapshots of the long-time spatial 
distribution of active disks in space for $a=1$.
(b) The corresponding  temporal evolution 
of the packing fraction starting 
with a random configuration of 50 disks of each type. 
The snapshot in
(a) is taken at the time indicated 
in the time series in (b).
The remaining parameters are  
$\beta_0 = 0.05,\delta=0.01,D=0.05,v=1$.}
\label{Fig:Snapshot_Dom}
\end{center}
\end{figure*}
\subsubsection{Setup and general behaviour}
We now study the third scenario, namely the case in which one type of particle always has a higher fitness than the other type, regardless of the number of interaction partners. Without loss of generality we focus on
the case where the type-$B$ particles
dominate. Specifically, we set  $c>0, d=c$ and $a=b=0$ so that 
\begin{equation}
\begin{aligned}
    \Pi_i^A&=0,\\
    \Pi_i^B&=c (N_i^A + N_i^B)\geq 0.
\label{matrix_dom}
\end{aligned}
\end{equation}
This means that the birth rate of a type-A particle  ($\beta^A=\beta_0$) is not affected by the number of particles in its vicinity. The birth rate for B-type particles is always higher than that for type-A, and is given by $\beta^B=\beta_0(1+ \tanh \Pi_i^{B})$. We have set $\alpha=1$.

In a first numerical experiment we start from a few particles of both types, randomly distributed in the system.  As time evolves both populations increase,  but at some moment, when the particle density is high 
enough, the population of type-$A$ particles starts
 decreasing, and eventually type A becomes extinct. An example is shown in Fig.~\ref{Fig:Snapshot_Dom} (b). A typical spatial configuration
at transient times can be seen in Fig.~\ref{Fig:Snapshot_Dom}(a). At difference with  the case of bistability, we do not find a transient
segregation pattern. 

In the next subsections we 
study the extinction
time and the fixation probability.

\subsubsection{Extinction time}
\label{sub:te_dom}

\begin{figure}[hbt!]
\begin{center}
\includegraphics[width=1.0\linewidth]{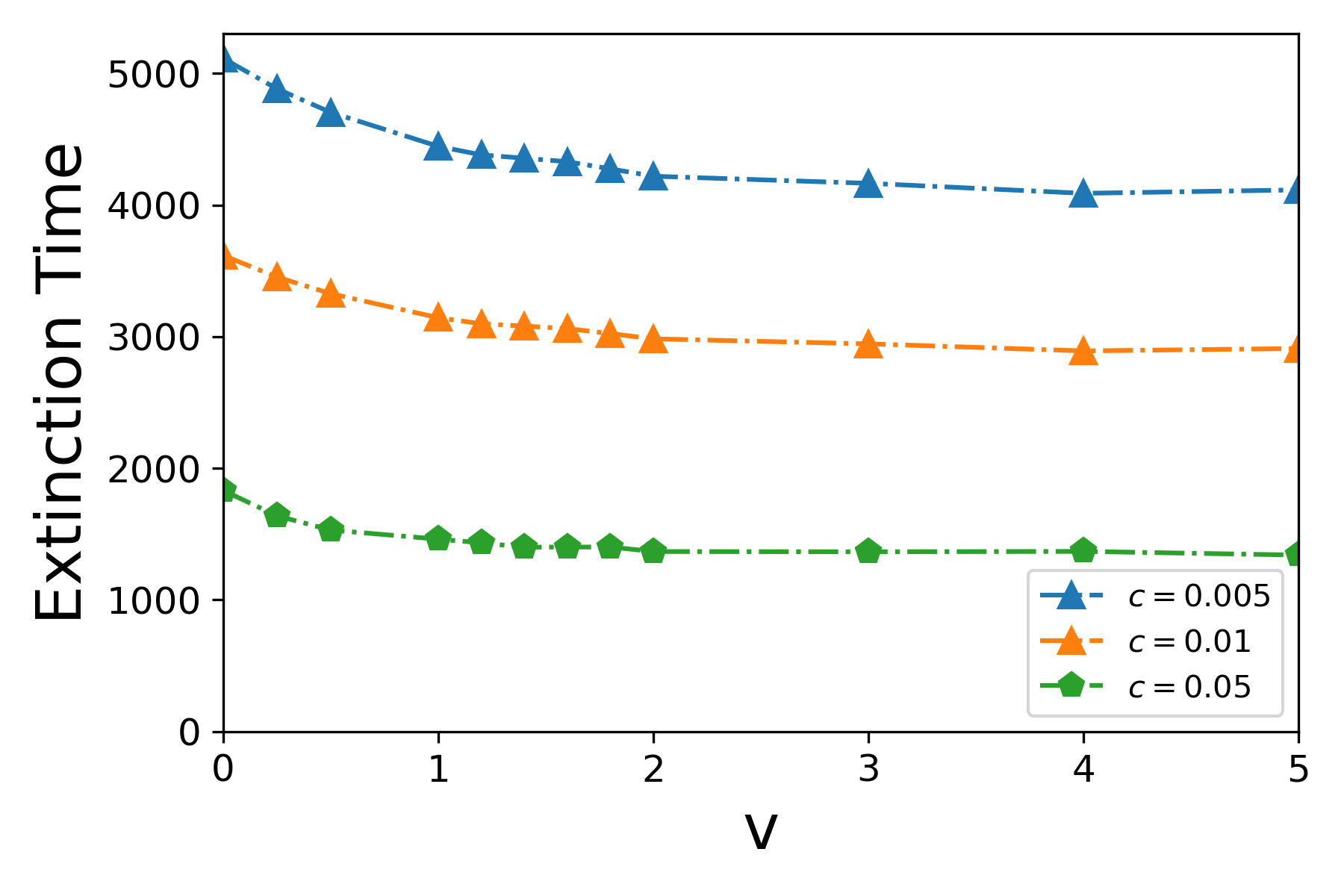}
\caption{Extinction time vs $v$ 
for different values of $c$. The remaining parameters 
are  $\beta_0 = 0.05,\delta=0.01,D=0.05$.}
\label{Fig:ext_time_dom}
\end{center}
\end{figure}

We first calculate the extinction time performing the numerical experiment indicated above for different choices of the activity $v$, and different values of the interaction parameter $c$ (see Fig.~\ref{Fig:ext_time_dom}). Increasing $c$ means to strengthen the extent by which the game dynamics favor particles of type B. By increasing $c$, the birth rate of type-B particles increases, and consequently, the time it takes for type A to go extinct reduces, as seen in Fig.~\ref{Fig:ext_time_dom}.

For any fixed value of $c$, the extinction time decreases as the mixing induced by higher activity $v$ increases. This occurs because particles with higher activity move rapidly through low-density regions but spend more time in high-density regions due to frequent collisions with other particles. These are also the areas where the birth rate is higher for type-B particles. However, while the extinction time decreases as $c$ increases, the difference in extinction time between low and high activity remains relatively small. This is because the birth rate of type-B particles depends on the local density, which is not significantly altered by increased activity. As a result, the parameter $c$, which controls the sensitivity of the birth rate to density, has a much stronger influence on the extinction time than the activity $v$.

\subsubsection{Fixation probability}
\label{sub:fp_dom}

We can obtain an analytical approximation for the probability
of dominance of the lineage of a single B-type particle invading the system. Broadly speaking this follows the methods of \cite{Traulsen2008}. If the system contains $m$ particles of type B at a given time, then it can transition to states $m\pm 1$. We can then formulate a relation between the dominance probabilities from these three states, $m$ and $m\pm 1$. This can then be solved recursively. The effective birth and death probabilities for these recursion relations can be estimated by assuming that the invading particles of type B live and die in an environment set by a stationary system of type-A particles only, details can be found Appendix \ref{ApB}. We find
\begin{equation}
\begin{aligned}
    P_{\rm B-domination} 
    &=\frac{ \tanh(\pi R^2 \rho c)}{1+ \tanh(\pi R^2 \rho c)}.
    \label{Eq:Domination_tanh}
\end{aligned}
\end{equation}
 We have assumed that the system is spatially homogeneous (which is a good approximation for small $v$)  with particle density
$\rho = N/L_s^2$. We introduce one particle of type B when the system of fully type A particles reaches the steady state. Therefore, the value of $N$ is the number of particles in this steady state. The quantity $\pi R^2 \rho $ is thus
the number of particles in disk of radius $R$. From the expression in Eq.~(\ref{Eq:Domination_tanh}) we find that $P_{\rm B-domination}=0.5$ in the limit of very large $c$. 

\begin{figure}[hbt!]
\begin{center}
\includegraphics[width=1.0\linewidth]{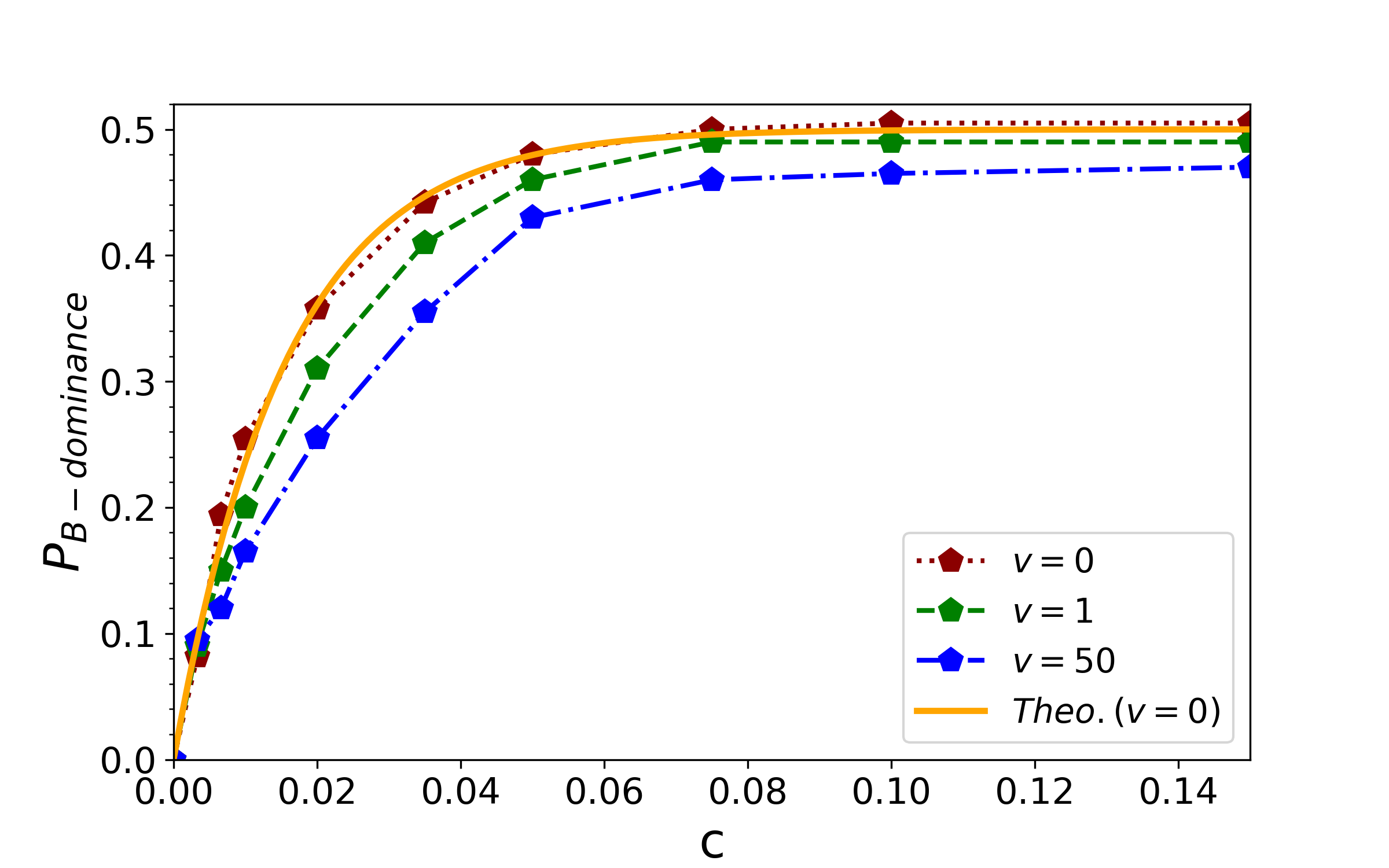}
\caption{Probability of dominance for the lineage of the invading type B as a function of the payoff parameter $c$, for 
different activities $v$.
Since we have $3200$ type-$A$ particles homogenously
distributed so that 
$\pi R^2 \rho \approx 32$,
thus
$P_{\rm B-domination}= 
\frac{\tanh(32 c)}{1+ \tanh(32 c)}$, which
is the solid line yellow line. 
The remaining parameters are  
$\beta_0 = 0.05,\delta=0.01,D=0.05$.}
\label{Fig:prob_dom}
\end{center}
\end{figure}

In Fig.~\ref{Fig:prob_dom}, we show the probability of dominance as a function of $c$ for different values of the activity $v$. This probability was determined by running multiple simulations with the same initial configurations and measuring the fraction of trials in which the introduced particle of the other type dominates the system. Our approximation becomes less accurate for large values of $v$, where the system becomes less homogeneous due to the emergence of the MIPS phase. We see that the probability of dominance in general decreases with the activity, and that even for very large values of $c$, it does not reach $50\%$ for large $v$.

\newpage
\section{Summary and discussions}
\label{Sec:summary}

We have explored the dynamics of 
two populations of motile, 
proliferating finite-size particles 
subject to interactions in underlying games. 
This is performed by considering
a  birth-death 
dynamics which is influenced 
by the local density 
of both populations, 
allowing us to simulate 
situations in which  
the growth rate of 
each population depends not only on
its own local density but also on the
density of the other type. 
This setup naturally 
leads to a variety 
of outcomes, including 
coexistence, bistability, 
and dominance, depending on
the payoff matrix of the game.

Coexistence occurs when both populations maintain stable proportions over long timescales. Our results show that increasing activity enhances spatial mixing by allowing particles to cross larger areas, and to find particles of the opposite type more frequently. In this scenario, what we observe is that for stronger interaction (high $|a|$ in our notation), the particles are more mixed, i.e., they are more likely to have balanced numbers of neighbours across the two types. In addition, stronger interaction promotes coexistence by amplifying the birth rate in regions with mixed particle types.

In the bistable scenario, the system evolves to a state dominated by one type. Which type this is depends on the initial conditions. Here, particles tend to cluster with those of the same type, leading to transient segregation and, eventually, the dominance of one type. The extinction time decreases with increasing diffusivity, or activity, as higher mobility promotes mixing and accelerates the elimination of one type. Interestingly, particles with higher diffusivity are more likely to dominate over others of low diffusion. In addition, stronger interactions, result in faster extinction due to the increased segregation.

Dominance arises when one type consistently has a higher payoff than the other, for given surroundings. Specifically, when we study the dominance scenario, we choose parameters such that type-B particles dominate the system due to their enhanced birth rate when surrounded by particles of any type, while type-A particles have a constant birth rate (independent of any other particles nearby). The extinction time decreases with increasing dominance (parametrized by $c$ in our setup), as stronger interactions accelerate the replacement of type-A particles, and with higher activity $v$, as particles explore the system more efficiently. However, the impact of activity on extinction time is limited compared to that of $c$, as the birth rate mainly depends on local density. Furthermore, the system’s resilience to the introduction of foreign particles increases with higher activity, as the probability that an introduced particle dominates the system decreases. This occurs due to reduced homogeneity at high activity levels, which limits the ability of the foreign particle to establish dominance. Even for large values of $c$, the dominance probability remains below $50\%$ when $v$ is high.

These findings emphasize the complex interplay between activity, interaction strength, and diffusivity in shaping the dynamics of proliferating motile-particle populations. They offer a framework for understanding how local interactions and mobility affect population outcomes in biological and ecological systems. To extend this work, future studies could incorporate more complex games with interactions between more than two particles at a time (i.e., multi-player games). One could also study the impact of varying particle sizes or motion types to simulate real-world systems more accurately, analyze systems with more than two species to understand multi-species competition and coexistence. Further work could focus on environmental factors such as gradients or dynamic boundaries. Our study provides insights into active matter systems with game-dynamical interactions and highlights the potential for modeling diverse biological and ecological phenomena.

\acknowledgments

A.A and C.L. acknowledge grant
LAMARCA PID2021-123352OB-C32 
funded by MCIN/AEI/10.13039/501100011033323
and FEDER “Una manera de hacer Europa”.
T.G. acknowleges
partial financial support from the Agencia Estatal
de Investigaci\'on and Fondo Europeo de Desarrollo 
Regional (FEDER, UE) under project APASOS 
(PID2021-122256NB-C21, PID2021-122256NB-C22), 
and the Maria de Maeztu programme for Units of Excellence, 
CEX2021-001164-M funded by  MCIN/AEI/10.13039/501100011033.

\appendix

\section{PROBABILITIES OF EXTINCTION IN THE STOCHASTIC SYSTEM}
\label{ApB}

In this appendix we derive an analytical approximation for the probability that the lineage of a single invador goes extinct \cite{harris1963theory, liggett1985, Nowak2006, Traulsen2008}. Our calculation applies to the scenario in Sec.~\ref{sub:Domination}, i.e., a situation in which selection favours the invading type B.

Initially the population is assumed to consist of $N$ particles of type A. We also assume that the system has a maximum capacity of $M$ particles, where $N, M \gg 1$. We then introduce one particle of type B. 

We use the following notation: $\phi_m$ represents the probability that the invader's lineage goes extinct if there are $m$ invaders in the system. We now look at the system starting from its current state. The next event changing the number of invaders is either a completed birth ($m\to m+1$), or a death event ($m\to m-1$). The probability for the former is written as $b$ and that of the latter is $1-b$. We assume that this quantity, $b$, remains constant. This can be justified as follows. If the invader's lineage goes extinct then it is fair to assume that the number of invaders will never become very high. Thus we can think of a small number of invaders in a `sea' of type-A particles. (We recall that we start the dynamics by inserting one type-B particle into the stationary state of a population made up entirely of type-A particles.) We also assert that the number of type-A particles and their distribution in space is not affected by the invaders. Type-B particles only see type A particles. The density of the latter in space determines $b$. Therefore, we can assume that $b$ remains approximately constant. Our calculation is for the scenario in which invader birth is more likely than invader death, i.e., $b>1/2$.

Due to the Markovian nature of the dynamics, then 
\begin{equation}
\phi_m=b\phi_{m+1}+(1-b)\phi_{m-1}.
\label{prob_ext}
\end{equation}
This is because, if there are currently $m$ invaders, then after the next event changing this number, the system is either in a state with $m-1$ or with $m+1$ invaders. These states occur with probabilities $b$ and $1-b$ respectively. Extinction then has to occur from this new state. 

Eq.~(\ref{prob_ext}) be re-arranged as follows,
\begin{equation}
\phi_{m+1}-\phi_m=\gamma (\phi_m-\phi_{m-1}),
\label{prob_ext_re}
\end{equation}
where $\gamma=(1-b)/b$. 

We now follow the steps in \cite{Traulsen2008}. We introduce $y_{m+1} = \phi_{m+1} - \phi_m$, and find from Eq. (\ref{prob_ext_re}),
\begin{equation}
y_{m+1} = \gamma y_m.
\label{prob_ext_y}
\end{equation}

Thus, we have for $m\geq 1$,  
\begin{equation}
y_m=\gamma^{m-1}y_1.
\label{prob_ext_y_re}
\end{equation}
Further,
\begin{align*}
    \sum_{m=1}^{M} y_m&=y_1+y_2+y_3 ...\\&=(\phi_1-\phi_0)+(\phi_2-\phi_1)+(\phi_3-\phi_2)+...\\
    &=-\phi_0+\phi_M.
\end{align*}

The quantity $\phi_m$ is the probability of extinction of invaders if there are $m$ invaders. Thus, $\phi_0=1$, and $\phi_M = 0$ (the latter indicates that extinction cannot occur if the system is entirely populated occupied by invaders, and if their number is large, $M\gg 1$). Then: 
\begin{align}
    \sum_{m=1}^{M} y_m=-\phi_0+\phi_M=-1.
\label{eq_l}
\end{align}

Using Eq.~(\ref{prob_ext_y_re}) on the other hand, we have
\begin{equation}
    \sum_{m=1}^{M} y_m=y_1 \sum_{m=1}^{M}  \gamma^{m-1}=y_1 \frac{1-\gamma^M}{1-\gamma}.
\end{equation}
We have $\gamma=(1-b)/b$, where $b>1/2$. Therefore, $\gamma<1$. We also assume $ M \gg 1$, and so we can neglect the term $\gamma^M \ll 1$ in Eq.~(\ref{eq_l}). We are then left with
\begin{align}
\sum_{m=1}^{M} y_m=\frac{y_1}{1-\gamma}.
\label{eq_r}
\end{align}

Thus, from Eq. (\ref{eq_l}) and Eq. (\ref{eq_r}),
\begin{equation}
    \frac{y_1}{1-\gamma}=-1.
\end{equation}

Recalling that $y_1=\phi_1-\phi_0=\phi_1-1$, we finally obtain 
\begin{equation}\label{eq:phi1}
\phi_1=\gamma=\frac{1-b}{b}.
\end{equation}
This is the probability that the lineage of a single invador of type B will go extinct eventually.

We now proceed to approximate the ratio $(1-b)/b$ in terms of the model parameters.  Suppose, birth events of invaders are triggered with rate $\beta_B$, and that invaders die with rate $\delta$. Assume also that an invader birth event, once triggered, completes with probability $p_{C, B}$.

We distinguish between two cases: (i) an invader birth event is completed {\it before} the next invader death event, (ii) an invader death occurs {\it before} the next completed invader birth event. The probability of the former is 
\begin{equation}\label{eq:b}
b=\frac{\beta_B p_{C, B}}{\delta+\beta_B p_{C, B}}.
\end{equation}
and the probability of the latter,
\begin{equation}\label{eq:one_m_b}
    1-b=\frac{\delta}{\delta+\beta_B p_{C, B}}.
    \end{equation}

Next we estimate the completion probability $p_{C, B}$. It is instructive to first look at a system without invaders (i.e., a population consisting only of particles of type A). In the steady state the rate of completed birth events (for particles of type A) is the same as the rate of death events (otherwise there would be a systematic change in particle numbers, i.e., the system would not be in the stationary state). We write this as
\begin{equation}
    \beta_A p_{C, A} = \delta,
    \end{equation}
   where $\beta_A$ is the rate with which birth events (for particles of type A) are triggered, $p_{C, A}$ is the probability that a triggered birth event completes, and $\delta$ is the common death rate for both types of particle.

Thus, we find
\begin{equation}\label{eq:p_complete} p_{C,A}=\frac{\delta}{\beta_A}.
\end{equation}
in the steady state of a system containing only type-A particles.

 In our system we choose the death rate $\delta$ to be same for both types of particle. However, the birth rate for the invading type of particle differs from that of the resident type. More specifically, the rates with with birth rates are {\it attempted} differ for the two types (as these rates are affected by payoff). Nonetheless, the probability of completing a birth event, once triggered, only depends on the local density of particles and on particle size. In our system both types of particle have the same size, and we proceed based on the assumption of a spatially homogeneous particle density. Therefore, the completion probability given a birth attempt is the same for both types of particle. Thus, we can assume
\begin{equation}
p_{C,B}=\frac{\delta}{\beta_A},
\end{equation}
where we stress that the right-hand side contains $\beta_A$.

We then have, using Eqs.~(\ref{eq:phi1}), (\ref{eq:b}) and (\ref{eq:one_m_b}),
\begin{eqnarray}
    \phi_1&=&\frac{1-b}{b} \nonumber \\
    &=& \frac{\delta}{\beta_B p_{C,B}} \nonumber \\
    &=& \frac{\beta_A}{\beta_B}.
\end{eqnarray}
 Finally, in the system in Sec.~\ref{sub:Domination} we have $\Pi^A=0$ and $\Pi^B = c (N_i^A+N_i^B)$. That is the payoff to particles of type A is zero no matter by how many particles they are surrounded, and the payoff to a focal particle of type B is c times the number of particles (of any type) in a radius $R$ around the focal particle.

 Using Eq.~(\ref{birth_rate}) we can therefore write
 \begin{equation}
     \beta_A=\beta_0,
     \end{equation}
and
\begin{equation}
    \beta_B=\beta_0 [1+\alpha \tanh(c\rho \pi R^2)],
\end{equation}
where $\rho$ is the particle density per unit area. This leads to Eq.~(\ref{Eq:Domination_tanh}), recalling that $P_{\rm B-domination}=1-\phi_1$.

\newpage

\bibliographystyle{apsrev4-2}
\bibliography{bib_.bib}

\end{document}